# Unidirectional spin Hall magnetoresistance in topological insulator/ferromagnetic layer heterostructures

Yang Lv[1†], James Kally[2†], Delin Zhang[1†], Joon Sue Lee[2], Mahdi Jamali[1], Nitin Samarth[2*] and Jian-Ping Wang[1*]

[1]Department of Electrical and Computer Engineering, University of Minnesota, Minneapolis, Minnesota 55455, USA

[2]Department of Physics, The Pennsylvania State University, University Park, Pennsylvania 16802, USA

[†]equal contribution

[*]JP.W. e-mail: jpwang@umn.edu; N.S. e-mail: nsamarth@psu.edu

**The large spin orbit coupling in topological insulators results in helical spin-textured Dirac surface states that are attractive for topological spintronics. These states generate an efficient spin-orbit torque on proximal magnetic moments at room temperature. However, memory or logic spin devices based upon such switching require a non-optimal three terminal geometry, with two terminals for the 'writing' current and one for 'reading' the state of the device. An alternative two terminal device geometry is now possible by exploiting the recent discovery of a unidirectional spin Hall magnetoresistance in heavy metal/ferromagnet bilayers and (at low temperature) in magnetically doped topological insulator heterostructures. We report the observation of unidirectional spin Hall magnetoresistance in a technologically relevant device geometry that combines a topological insulator with a conventional ferromagnetic metal. Our devices show a figure-of-merit (magnetoresistance per current density per total resistance) that is comparable to the highest reported values in all-metal Ta/Co bilayers.**

The spin Hall effect (SHE) in non-magnetic (NM) heavy metals originates in their strong spin-orbit coupling (SOC) and has been extensively studied recently[1–4]. When a charge current flows through a NM heavy metal, the SHE yields a spin accumulation at the interface with a proximal material. If the latter is a ferromagnetic (FM) layer, the spin accumulation at the interface can exchange angular momentum with the magnetic moments and exert a spin-orbit torque (SOT). In certain configurations and at sufficiently high charge current density, the magnetization in the FM can be switched. SOT switching is believed to be potentially faster and more efficient than spin transfer torque (STT) switching that is typically used in magnetic tunneling junction (MTJ) devices for memory and logic applications[3,5,6].

SOT switching devices consist of a current carrying channel with a proximal nanomagnet whose magnetization determines the memory or logic state. Such devices need two terminals for 'writing' the state of the device and an additional terminal, usually an MTJ on top of the nanomagnet, for 'reading' the magnetization state of the device[3,5]. Since the stable states of the nanomagnet are 180-degree-opposite to each other, symmetry prevents the sensing of the magnetization state using a conventional two terminal magnetoresistance, such as anisotropic magnetoresistance (AMR) or spin Hall magnetoresistance (SMR)[7,8]. The required presence of a third terminal for reading makes such SOT switching devices more difficult to fabricate and usually less appealing for memory and logic applications.

With the recent discovery of unidirectional spin Hall magnetoresistance (USMR) in NM/FM bilayers, such as Pt/Co and Ta/Co, the third terminal of SOT switching devices is no longer necessary[9–11]. USMR originates from the interactions between the spins generated at the NM-FM interface by SOC of the NM and the conduction channels in the FM. The unique feature of USMR is its symmetry; it is sensitive to two opposite magnetization states. Therefore, this allows one to envision a two terminal SOT switching device that relies on USMR: the nanomagnet is switched by a current through the NM channel, while the state of the magnetization of the nanomagnet is simply read out using the USMR.

While much of the mainstream activity in SOT devices has focused on heavy metals, such as Ta, Pt and W, recent research has begun to explore the potential of 3D topological insulators (TIs) [12–15]. These are narrow band gap semiconductors wherein strong SOC and time-reversal symmetry yield helical spin-textured Dirac surface states whose spin and momentum are orthogonal. This 'spin-momentum locking' (SML) has been confirmed using direct measurements such as photoemission[13], electrical transport[16–19] and spin torque ferromagnetic resonance[20], as well as indirect means such as spin pumping[21–26]. It has also been demonstrated that the spins can exert torques on a FM[20,27] as one would expect of SOT in the NM/FM case.

In comparison to the NM/FM bilayers, where SOT switching and sensing using USMR have both been confirmed, the observation of USMR in TI/FM systems is just beginning to emerge. In a very recent study [28], a large USMR was observed in $Cr_x(Bi_{1-y}Sb_y)_{2-x}Te_3/(Bi_{1-y}Sb_y)_2Te_3$ bilayer structures at very low temperatures. Here, the Cr-doped layer is a FM TI with a low Curie temperature and the other layer is a NM TI. For more pragmatic applications, it is desirable to explore the USMR phenomenon in heterostructures that interface a TI with a conventional FM of technological relevance. Here, we report the experimental observation of USMR in TI/FM heterostructures, including $(Bi,Sb)_2Te_3$/CoFeB and $Bi_2Se_3$/CoFeB bilayers. As illustrated in Fig. 1, spins are generated due to the SML of the TI when a charge current, j, is applied in the bilayer. Depending on the relative directions between the spins and magnetization of FM, spins at the interface present different conductance when interacting with the conduction channels in the FM. The USMR in TI/FM systems is like that in NM/FM systems with the different mechanisms of spin generation.

We observed USMR at temperatures between 20 K and 150 K for $(Bi,Sb)_2Te_3$ (BST) and $Bi_2Se_3$ (BS). The largest USMR among our samples is about 2.7 times as large as the best USMR in Ta/Co sample, in terms of USMR per total resistance per current density, observed in 6 QL BS and CoFeB of 5 nm bilayer.

The devices studied are fabricated from BST ($t$ QL)/CoFeB (5)/MgO (2) and BS ($t$ QL)/CoFeB (5)/MgO (2) thin film stacks ($t$ = 6 and 10), grown by molecular beam epitaxy (MBE) and magnetron sputtering. Hall bars of nominal length 50 μm and width 20 μm are tested with harmonic measurements under both longitudinal and transverse resistance setup. The magnetization of CoFeB is spontaneously in-plane with little perpendicular anisotropy field.

Figure 2a shows the definition of the coordinates and rotation planes. Zero angles are at $x+$, $y+$ and $z+$ directions for $xy$, $zx$ and $zy$ rotations respectively. The directions for rotation for increasing angle are indicated by arrows. A 3 Tesla external field is applied and rotated in the $xy$, $zx$ and $zy$ device planes while the first order resistance $R_\omega$ and second order resistance $R_{2\omega}$ are recorded with 2 mA R.M.S. AC current. Figure 2b and 2c show the angle dependencies of $R_\omega$ and $R_{2\omega}$, respectively, of the BST (10 QL)/CoFeB (5 nm)/MgO (2 nm) sample at 150 K. The $R_\omega$ exhibits typical SMR-like behavior with the $R^x > R^z > R^y$. Similar to the behavior seen in all metallic NM/FM bilayers, the variation of the second order resistance $R_{2\omega}$ with angle is also proportional to the magnetization projected along the $y$-direction. The period of $xy$ and $zy$ rotations are 360 degrees while a flat line is observed in the $zx$ rotation. The amplitude of $R_{2\omega}$ is about 3 mΩ with an average current density of 0.667 MA/cm$^2$.

Due to Joule heating of the device and the temperature gradient across the device plane, the anomalous Nernst effect (ANE) and spin Seebeck effect (SSE) also contribute to the second order resistance. To carefully separate this contribution (denoted as $R_{2\omega}^{\Delta T}$,) from the USMR, we carried out a series of measurements of Hall or transverse second order resistance with $xy$-plane rotations under various external field strengths. Figure 3a shows the Hall resistance setup. The transverse resistance is measured while the external field is rotated in the $xy$-plane. The second order Hall resistance, $R_{2\omega}^H$, contains contributions from ANE/SSE, field-like (FL) SOT and anti-damping (AD) SOT. The ANE/SSE and AD SOT are proportional to $cos\varphi$ while the FL SOT is proportional to $cos3\varphi + cos\varphi$ (ref [29]). Fig. 3b shows two examples of $R_{2\omega}^{\Delta T}$ vs. angle with 20 mT and 3 T external fields, respectively. Since AD SOT and FL SOT perturb the magnetization and thus contribute to $R_{2\omega}^H$, their effects diminish at larger external field. Figure 3b shows that the data measured in a 20 mT field contain both $cos\varphi$ and $cos3\varphi$ components, while in a 3 T field, the data exhibit almost no $cos3\varphi$ component. There are two steps to obtain the $R_{2\omega}^{\Delta T}$. First, by fitting the angle dependent data, we extract the amplitudes of the $cos\varphi$ and $cos3\varphi$ components. The FL SOT can then be easily determined and separated. This leaves the ANE/SSE and AD SOT. We plot the data corresponding to these contributions versus the reciprocal of total field, as shown in Fig. 3c. In this figure, $B_{dem}$-$B_{ani}$ is the demagnetization field minus the perpendicular anisotropic field of the FM layer, which is determined to be about 1.5 T by separate measurements. Since the effect of the AD SOT will diminish at infinite field, the intercept of the fitted line is the contribution of ANE/SSE to the 2$^{nd}$ order Hall resistance. Then, we can obtain the contribution of ANE/SSE to the longitudinal resistance $R_{2\omega}$ by scaling that from the Hall resistance with the relative ratio of device length to device width. Finally, the USMR is determined once the ANE/SSE contribution is subtracted from the $R_{2\omega}$.

Figures 4a and b show the $R_{2\omega}$, $R_{2\omega}^{\Delta T}$ and $R_{USMR}$ of 10 QL BST (4a) and 10 QL BS (4b) samples with 2 mA and 3 mA current, respectively, at various temperatures. Temperature affects the chemical potential and the relative contributions to transport from surface and bulk conduction. As a result, even though the magnetization and resistivity of the CoFeB layer vary little within the range of temperature in our experiments, the charge to spin conversion in TIs and the related USMR are both strongly temperature dependent. The 10 QL BST/CoFeB sample gives its highest USMR at 70 K while the $R_{2\omega}$ and $R_{2\omega}^{\Delta T}$ keep increasing with increasing temperature up to 150 K. The USMR of 10 QL BS/CoFeB can only be confirmed within between 50 K and 70 K because of larger noise and magnetic field dependent signal outside the temperature window. At 70 K, BST and BS samples show resistance $R^z$ of 733 Ω and 488 Ω, and USMR per current density of 1.067 mΩ/MA/cm$^2$ and 0.633 mΩ/MA/cm$^2$, respectively. The ratios of USMR per current density to total resistance of the two samples are 1.45 ppm/(MA/cm$^2$) and 1.30 ppm/(MA/cm$^2$), respectively. These values are slightly better than the best result obtained using Ta/Co bilayers (1.14 ppm/(MA/cm$^2$) at room temperature)[10].

Figure 5 shows USMR per current density per total resistance ($R_{USMR}$/j/R) and sheet USMR per current density ($\Delta R_{USMR}$/j) of all four samples as a function of temperature (BS{x} or BST{x} are abbreviations of BS or BST samples of {x} QL thicknesses). This provides a more meaningful figure-of-merit for comparisons of USMR across different types of samples. These two values also show very similar trends for all samples at various temperatures, except for the comparison between BST6 and BS10 at 70K, in which BST6 is lower than BS10 in terms of $R_{USMR}$/j/R but higher in terms of $\Delta R_{USMR}$/j. The swap of position is mostly due to the larger total resistance of BS10 compared to BST6 while they show similar $R_{USMR}$/j. The largest $R_{USMR}$/j/R and $\Delta R_{USMR}$/j are 3.19 ppm/MA•cm$^2$ and 0.95 mΩ/MA•cm$^2$, respectively, and both observed in BS6 at 150K. It is more than as twice large as the best reported Ta/Co case[10]. The USMR measurements beyond the temperature ranges of the plots of each sample show strong noise and field-dependent signal background as to render the estimations of USMR unreliable.

In summary, we demonstrated the presence of USMR in topological insulator/ferromagnetic layer heterostructures. The USMR was observable with a much lower current density compared to all metallic NM/FM bilayers. The ratios of USMR per current density to total resistance are found to be comparable to the best result reported so far in Ta/Co bilayers. The observation of USMR in a TI/FM system is the last missing piece of the puzzle to build a two terminal TI-based SOT switching device. Such a two-terminal topological spintronic switching device is potentially more efficient compared to MTJs that use STT switching due to the large SOC of TIs. The USMR we observe will enable the read operation of such a device without having to build a MTJ structure on top of TI. Such two terminal devices are much more architecture friendly and easier embedded in current STT magnetic random access memory architectures.

**Methods**

The $Bi_2Se_3$ or $(Bi_{1-x}Sb_x)_2Te_3$ films were grown by MBE on InP (111) substrates. The InP (111) substrate is initially desorbed at 450°C in an EPI (Veeco) 930 MBE under high purity (7N) $As_4$ supplied by a Knudsen cell until a 2x2 reconstruction is visible in reflection high energy electron diffraction. The substrate is then moved under vacuum to an EPI 620 MBE for the Bi-chalcogenide deposition. $Bi_2Se_3$ films were grown from high purity (5N) Bi and Se evaporated from Knudsen cells at a beam equivalent pressure flux ratio of 1:14. The substrate temperature was 325°C (pyrometer reading of 250°C) and the growth rate was 0.17 nm/min. The films have a root mean squared (RMS) roughness of approximately 0.7 nm over a 25 $\mu m^2$ area measured by atomic force microscopy (AFM). For $(Bi,Sb)_2Te_3$ films, the flux ratio of Bi to Sb was 1:3 and (Bi+Sb):Te is at a flux ratio of approximately 1:12 for a growth rate of 0.44 nm/min with a RMS roughness of approximately 1.1 nm over a 25 $\mu m^2$ area measured by AFM. These films are grown at a substrate temperature of 315°C (240°C measured by a pyrometer) using 5N purity Sb and 6N Te from Knudsen cells. Film thickness is measured by X-ray reflectivity and crystal quality by high-resolution X-ray diffraction rocking curves of the (006) crystal plane—with a full width half max (FWHM) of approximately 0.28 and 0.11 degrees for $Bi_2Se_3$ and $(Bi,Sb)_2Te_3$ films respectively.

The MBE-grown TIs were then sealed in Argon gas and transported to an ultra-high vacuum (UHV) six-target Shamrock sputtering system which could achieve a based pressure better than $5\times10^{-8}$ Torr at room temperature. The thin films were first gently etched by Argon ion milling. Then the CoFeB layer was deposited using a $Co_{20}Fe_{60}B_{20}$ target. Finally, an MgO layer was deposited to serve as a protection layer.

The device fabrication began with a photolithography followed by an ion milling etching to define the Hall bars. Then the second photolithography and an e-beam evaporation followed by a liftoff were performed to make contacts.

The devices were tested in a Quantum Design PPMS which provides temperature control, external field and rotation. The AC current at 10 Hz was supplied by a Keithley 6221 current source. A Stanford Research SR830 or an EG&G 7265 lock-in amplifier paired with an EG&G 7260 lock-in amplifier were used to measure the first and second harmonic voltages, respectively and simultaneously.

**Acknowledgements**

This work was supported in part by C-SPIN, one of six centers of STARnet, a Semiconductor Research Corporation program, sponsored by MARCO and DARPA. We would like to thank Timothy Peterson and Gordon Stecklein for their help on the usage of PPMS.

**Author Contributions**

YL and JPW conceived and designed the experiments. JSL, JK and NS grew and provided TI thin films. DZ and MJ deposited the rest layers of the thin film stacks. YL fabricated and tested devices and processed data. All authors reviewed results. YL and JPW wrote the manuscript. All authors contributed to the completion of the manuscript.

**Competing Financial Interests Statements**

The authors declare no competing financial interests.

(2014).

**Figure Legends**

Fig. 1. Illustration of USMR in TI/FM bilayer. Spins are generated at interface when a charge current is applied. The relative directions of spins to magnetization of either parallel a) and anti-parallel b) result in different resistance states.

Fig. 2. a) Longitudinal resistance measurement setup and definitions of rotation planes. First order b) and second order c) resistances of 10 QL BST sample at 150 K are shown when the external field is rotated in three orthogonal planes. The starting points and zero angles are at $x+$, $y+$ and $z+$, the directions of rotation of increasing angle are $x$ to $y$, $z$ to $x$ and $z$ to $y$, for $xy$, $zx$ and $zy$ rotations, respectively.

Fig. 3. a) Transverse/Hall resistance measurement setup; b) Examples of second order Hall resistance of 10 QL BST sample at 150 K vs. angle in $xy$ plane rotation with 20 mT and 3 T external fields; c) Hall resistance measured with various external fields are plotted vs. reciprocal of total field and linear fitted. The intercept of the fitted line represents the contribution of ANE/SSE.

Fig. 4. The measured second order longitudinal resistance $R_{2\omega}$ consists of contribution of $R_{2\omega}^{\Delta T}$ and USMR $R_{USMR}$. Each component is plotted vs. temperature for a) 10 QL BST sample and b) 10 QL BS sample.

Fig. 5. a) USMR per current density per total resistance and b) sheet USMR per current density of all four samples at various temperatures. BS{x} or BST{x} are abbreviations of BS or BST samples of {x} QL thicknesses.

**Figures**

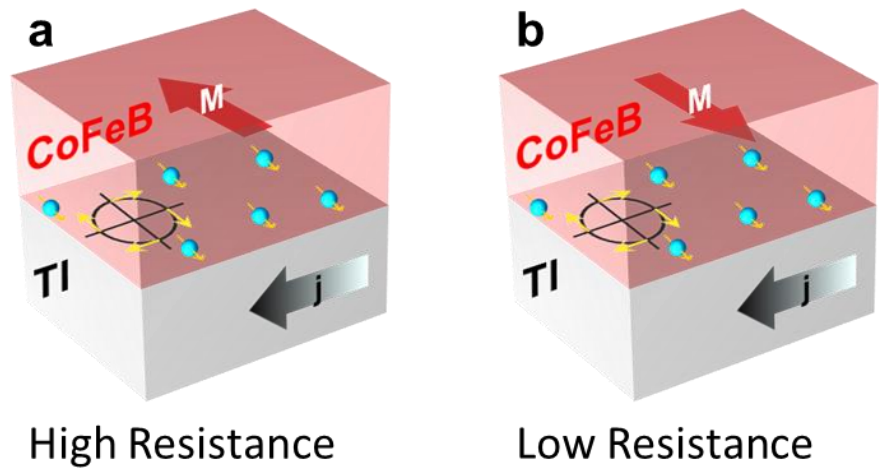
8

Fig. 1

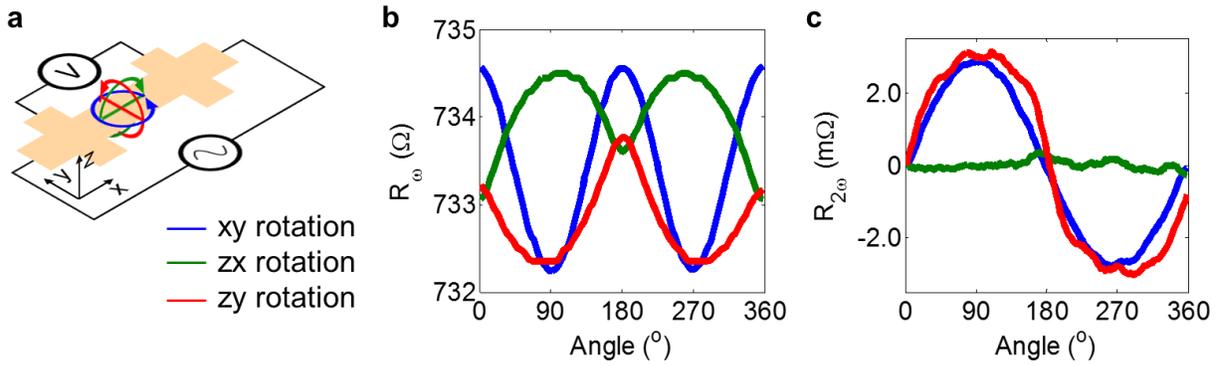

Fig. 2

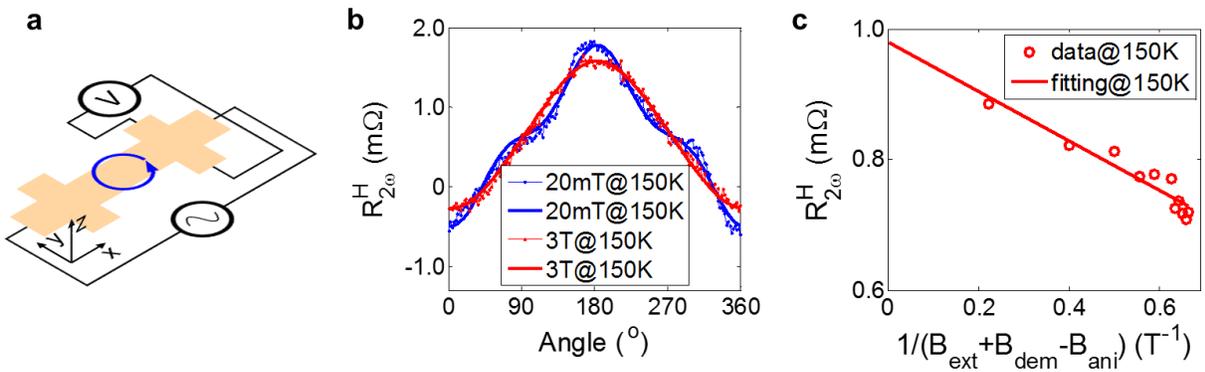

Fig. 3

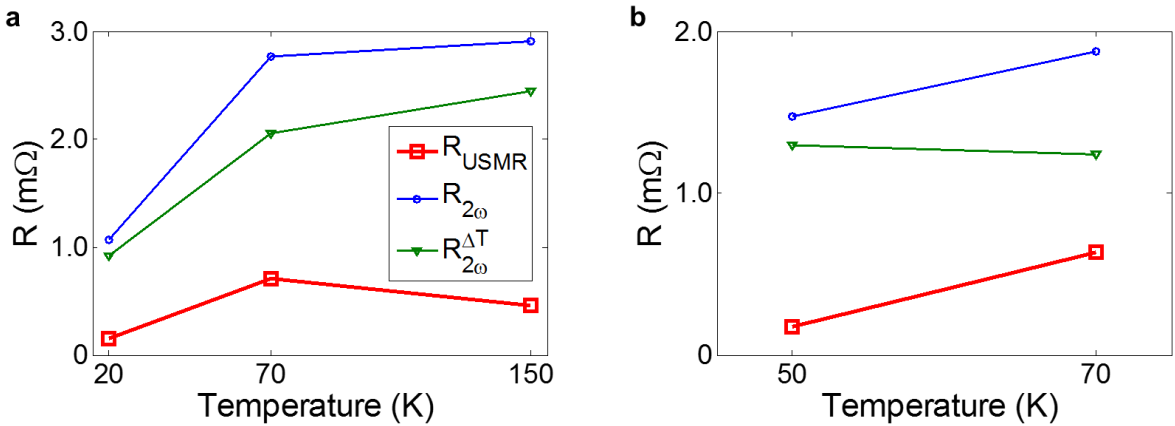

Fig. 4

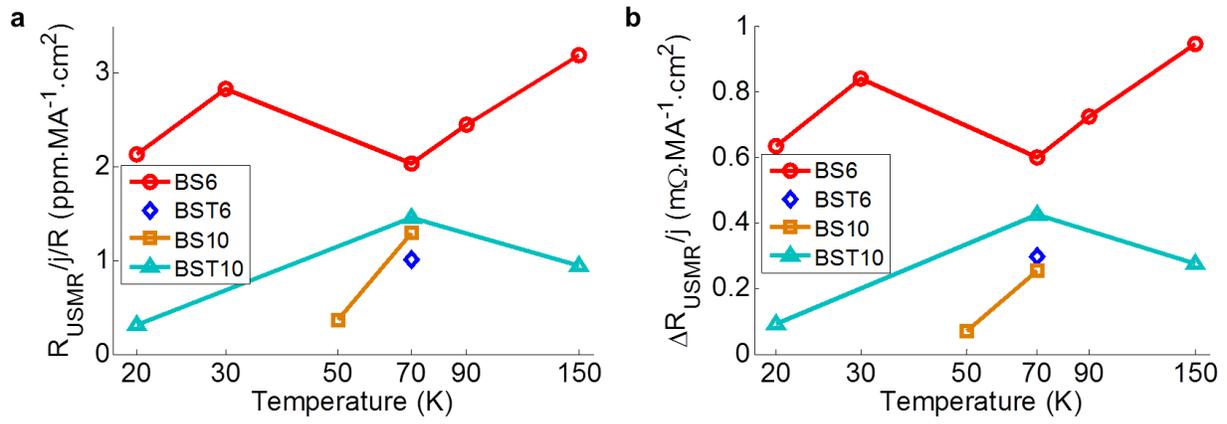

Fig. 5